\documentclass{article}

\usepackage[final]{nips_2018}

\usepackage[utf8]{inputenc} 
\usepackage[T1]{fontenc}    
\usepackage{hyperref}       
\usepackage{url}            
\usepackage{booktabs}       
\usepackage{amsfonts}       
\usepackage{nicefrac}       
\usepackage{microtype}      

\title{Scalable Graph Learning for Anti-Money Laundering: A First Look}

\author{
  Mark~Weber \\
  MIT-IBM Watson AI Lab\\
  IBM Research\\
  \texttt{m.weber@sloan.mit.edu} \\
  \And
  Jie Chen \\
  MIT-IBM Watson AI Lab\\
  IBM Research\\
  \texttt{chenjie@us.ibm.com} \\
  \And
  Toyotaro Suzumura \\
  MIT-IBM Watson AI Lab\\
  IBM Research\\
  \texttt{tsuzumura@us.ibm.com} \\
  \And
  Aldo Pareja \\
  MIT-IBM Watson AI Lab\\
  IBM Research\\
  \texttt{aldo.pareja@ibm.com} \\
  \And
  Tengfei Ma \\
  MIT-IBM Watson AI Lab\\
  IBM Research\\
  \texttt{tengfei.ma1@ibm.com} \\
  \And
  Hiroki Kanezashi \\
  MIT-IBM Watson AI Lab\\
  IBM Research\\
  \texttt{hirokik@us.ibm.com} \\
  \And
  Tim Kaler \\
  MIT-IBM Watson AI Lab\\
  MIT CSAIL\\
  \texttt{tfk@mit.edu} \\
  \And
  Charles E. Leiserson\\
  MIT-IBM Watson AI Lab\\
  MIT CSAIL\\
  \texttt{cel@mit.edu}\\
  \And
  Tao B. Schardl\\
  MIT-IBM Watson AI Lab\\
  MIT CSAIL\\
  \texttt{neboat@mit.edu}\\
}

\begin{document}

\maketitle

\begin{abstract}
Organized crime inflicts human suffering on a genocidal scale: the Mexican drug cartels have murdered 150,000 people since 2006; upwards of 700,000 people per year are ``exported'' in a human trafficking industry enslaving an estimated 40 million people. These nefarious industries rely on sophisticated money laundering schemes to operate. Despite tremendous resources dedicated to anti-money laundering (AML) only a tiny fraction of illicit activity is prevented. The research community can help. In this brief paper, we map the structural and behavioral dynamics driving the technical challenge. We review AML methods, current and emergent. We provide a first look at scalable graph convolutional neural networks for forensic analysis of financial data, which is massive, dense, and dynamic. We report preliminary experimental results using a large synthetic graph (1M nodes, 9M edges) generated by a data simulator we created called AMLSim. We consider opportunities for high performance efficiency, in terms of computation and memory, and we share results from a simple graph compression experiment. Our results support our working hypothesis that graph deep learning for AML bears great promise in the fight against criminal financial activity.






\end{abstract}

\section{Anti-Money Laundering in 2018}

Anti-money laundering (AML) is the task of preventing criminals from moving illicit funds through the financial system. AML is generally viewed through the lens of regulatory compliance because the burden of forensic analysis falls primarily on financial institutions, which are responsible for meeting Know Your Customer (KYC) standards, monitoring transactions, shutting down or restricting accounts deemed suspect, and submitting timely Suspicious Activities Reports (SARs) to law enforcement agencies. In the United States, these requirements are primarily stipulated in the Bank Secrecy Act (1970) and the Patriot Act (2001), with enforcement chiefly by the Financial Crimes Enforcement Network (FinCEN). Global costs of AML compliance run in the tens of billions of dollars and have been growing at about 15\% per year since 2004 \citep{KPMG2014}. Penalties for failed AML compliance are severe: in 2018, the Commonwealth Bank of Australia unhappily set a new record with a USD \$530 million fine for AML violations involving its ``Intelligent Deposit Machines'' (a cautionary reminder that calling things ``intelligent'' does not make them so) \citep{BBC2018}. 

Meanwhile we are rapidly becoming a cashless society engaged in global economic exchange. The advent of cryptocurrency has catalyzed a paradigm shift in peer-to-peer transactions and extranational financial governance. Cryptocurrencies present new challenges but also new opportunities for AML, and these vary across type (e.g. forensic insight into Bitcoin's public ledger is much better than trying to track traditional cash, but more privacy-preserving cryptocurrencies like Zcash are far more difficult to monitor).

Thus, we observe two rival, converging forces: 1) the increasing difficulty of AML in the age of globalization and cryptocurrency; 2) the acceleration of advanced forensic tools in the age of big data and artificial intelligence. Our aim is to ensure the second outpaces the first.

\subsection{A pressing societal problem}
Amidst the legal and econometric detail, the real societal problem too often fades into the background: tens of millions of innocent people are suffering under the malice of drug cartels, human trafficking rings, and terrorist organizations. Moreover, the resulting AML efforts of governments become a force for financial exclusion as financial institutions pass along high compliance costs to customers and turn more people away; low-income people, immigrants, and refugees are hit the hardest \citep{WorldBank2012}. 

\subsection{The bad guys are winning}

Today, we are woefully ineffective at preventing money laundering. Europol estimates only 1\% of illicit funds are confiscated \citep{Europol2017}. To understand why, we consider both the technical and behavioral challenges, as well as the current methods for addressing them.

Technically, we have a needle-in-a-haystack problem of entity classification and hidden pattern discovery in massive, dynamic, high dimensional, time-series transaction data sets with high noise-to-signal ratios, combinatorial complexity, and non-linearity. Data sets are often fragmented, inaccurate, incomplete, and/or inconsistent, within as well as across organizations. Synthesizing information from multimodal data streams is difficult to automate, and so it falls to resource-constrained human analysts. Behaviorally, the heavy hand of compliance (i.e. the cost asymmetry of a false positive versus a false negative) motivates over-reporting, leading to a secondary needle-in-a-haystack problem for resource-constrained government law enforcement.

A recent Reuters survey of c-suite executives at 2,373 large global organizations provides color on how these challenges play out for stakeholders. An estimated 5\% of reviewed transactions are filed as Suspicious Activities Reports (SARs) and only 10\% of those receive a meaningful investigation by law enforcement. Thus 0.5\% of criminal activity alerts lead to action. Each organization transacts with thousands of third-party vendors and millions of clients, with finance and insurance companies comprising the upper fractiles of the distribution. Internal monitoring is just as important; respondents estimated 58\% of financial crimes involving human trafficking were internal to their own organizations \citep{Reuters2018}.

\section{Review of current methods}

AML procedure follows five key steps: (1) installation of a compliance organization within the company including formal AML training for employees, often with external support from specialized firms such as Promontory; (2) development and execution of Know Your Customer (KYC) on-boarding and profile maintenance procedures; (3) account activity oversight and constraints via transaction monitoring systems (TMS); (4) manual review of flagged accounts and transactions; (5) filing of Suspicious Activities Reports (SARs) to law enforcement and corresponding restrictive action against suspect accounts. We consider Steps 2 and 3: transaction monitoring and evaluation.

\subsection{Transaction Monitoring}

Transaction monitoring systems are predominantly rules-based thresholding protocols tuned for volume and velocity of transactions with tiered escalation procedures. For example, a U.S. bank might outsource Level 1 (our term) transaction monitoring to a company in India, which flags all transactions above \$10,000 for review, including ``near misses'' such as a transaction of \$9,999 and/or transaction series such as five transactions of \$2,000 each in a 24 hour window. A Level 2 analyst performs a preliminary review of each flag and decides whether or not to escalate to the bank's headquarters back in the U.S. There, a Level 3 analyst, in coordination with the compliance team, will conduct a more thorough analysis using sophisticated techniques to determine whether or not a SAR need be filed and the account in question suspended. 

\subsection{Forensic Analysis}

\textbf{Suspiciousness heuristics} govern the initial review of a flagged account. These include political exposure, geographic dynamics, round numbers, transaction type and properties (e.g. if the transaction is a purchase of an asset, how liquid is the asset and how stable is it as a store of value?), and behavioral logic (e.g., something does not add up). 

For example, ``Account A'' is flagged for a transaction of \$9,500, a ``near-miss'' on the \$10,000 rule. The Level 2 analyst pulls up the KYC profile and sees the account belongs to U.S. senator (a Politically Exposed Person or ``PEP'' in AML parlance) has been credited by a foreign account in the Bahamas, a money-laundering hotspot. The analyst pulls up more transaction activity and also sees a \$4,000 credit from XYZ Corporation, which the analyst cannot find record of in a quick Google search (why is a PEP on a mystery corporate's payroll?). The Level 2 analyst deems this transaction/account suspicious and escalates to the Level 3 analyst for further review.

In this toy scenario, suspiciousness is obvious; reality is far more difficult. First, the aforementioned needle-in-the-haystack problem, compounded by the threat of fines, produces high false positive rates. Second, the manual suspiciousness heuristics above do not scale well with time and human resource constraints (analysts are often significantly backlogged). Third, criminals can be quite sophisticated in masking the true nature of their transactions with complicated account layering, multi-hop transactions, and other techniques. More sophisticated AML methods are needed to meet these challenges.

\textbf{Graph analytics} have emerged as an increasingly important tool for AML analysis because money laundering involves cash flow relationships between entities i.e. network structures. Multiple graph types can be constructed. For example, a graph can be formulated where a single account is represented as a vertex and a single transaction between two accounts is represented as an edge. Another approach is to represent a group of accounts as a vertex (e.g. such as those under a holding company, or such as those inferred to share an owner via clustering) and define an edge as the aggregate transaction volume with a neighboring node over a period of time. The latter is the predominant method for forensic analysis of cryptocurrency activity, involving a combination of 1) address clustering using multiple-input and change heuristics, and 2) address tagging with attribution data gathered from public sources \citep{Ransomware}. In the traditional financial system, where KYC rules assist with entity identification, classical graph algorithms such as cycle detection, PageRank, degree distribution, egonet, and label propagation are used to identify anomalies, relationships, and other features of the network that may be deemed suspicious. Working with private bank data, we have found these methods can reduce false positives by 20--30\% when performed by expert analysts.

\textbf{Natural Language Processing} (NLP) is a newer tool with emergent potential for AML. \cite{NextGenAML2018} recently demonstrated how NLP processing of unstructured, heterogeneous data from online sources can be used in combination with transaction monitoring systems and private bank data to produce suspiciousness scores and visualizations to augment human forensic analysis. In this preliminary framework, the authors tap open data streams such as news articles, social media, financial reports, and fraud databases. Then, cross-referencing with the private bank data such as KYC and transaction activity, they use a sequential combination of sentiment analysis, name entity recognition, and relation extraction to build knowledge graphs and relation graphs pertinent to the flagged account. The authors report positive stakeholder feedback on their stated objective to reduce average investigation times by 30\%. We find this work compelling, but in need of more powerful and efficient graph methods to support the effective and efficient analysis of such graphs at scale.

\textbf{Graph learning} using deep neural networks is a new method with great promise for the complex pattern and feature discovery central to the AML challenge, as well as for the aforementioned multimodal constructions of various graph types. We explore this in more detail in the next section.

\section{Deep learning for massive, dynamic graph data}

The proliferation of graph analytics in AML illuminates the potential for ever more powerful graph methods such as graph learning, an emerging sub-field of machine learning.

\subsection{Scalable Graph Convolutional Networks (GCN)}

Deep learning has achieved remarkable success on Euclidean data, including images, audio, and video. Deep learning for graph structured data, however, has lagged due to the scalability challenges of dealing with combinatorial complexity and non-linearity inherent to graphs of any meaningful size and density. But this is changing.


Recently, a proliferation of graph neural network models have emerged~\citep{Bruna2014,Defferrard2016,Li2016,Hamilton2017,Gilmer2017}, and several have made major strides toward scalability with promising experimental efficiency results \citep{Chen2018,Ying2018}. We prioritize FastGCN~\citep{Chen2018} in the next section, as it was demonstrated on benchmark data sets to surpass peer methods GCN and GraphSAGE by as much as two orders of magnitude without sacrificing accuracy.

\subsection{Graph learning for AML}

The challenge of AML involves high dimensional, massive graph data mapping billions of relationships (edges) between millions of entities (nodes). In AML transaction monitoring, a node entity might be a single account or a set of associated accounts (either known or inferred via clustering to operate as one). A node could also represent another graph from a previous step in a time series. 

The known properties are those explicitly defined in the data, such as information collected in standard Know Your Customer (KYC) processes or multi-modal data mined from public or partner information streams, as well as the observable transactions and any associated flags or SARs filed. The unknown, hidden properties (such as covert criminal relationships and activity) are the subject of the forensic inquiry; i.e., what more can we ascertain about an entity than the human eye or classical graph analytics can detect? In the language of deep learning, we begin with an initial, often quite limited, attribute vector of a node entity, and we aim to arrive at a much more characteristic feature representation with a higher degree of confidence.

The principle challenge is scalability as an inverse function of graph size, and in AML the graph size is massive and aways increasing (new accounts and interactions are added constantly, producing an ever-growing graph, or graphs within graphs).

\subsection{Preliminary testing of graph convolutional networks for AML}

We have conducted preliminary experiments using graph learning models for AML. We detail one such experiment here:

\textbf{Synthetic data:} Since transaction data and customer profiles are privacy sensitive, we created a prototype data simulator called AMLSim (to be described in detail in a forthcoming paper). AMLSim is a multi-agent simulation platform tailored for an AML problem, where each agent behaves as a bank account transferring money to other agent accounts, and where a small number of agents conduct nefarious activity modeled on real-world patterns. AMLSim constructs data in two steps: (1) generate a graph via NetworkX\footnote{https://networkx.github.io} based on a given degree distribution; (2) generate a time-series of transactions via PaySim~\citep{Lopez2016}, based on transaction distributions and dynamics observed in real data. We draw on domain expertise to define for the simulator a list of known suspicious patterns. This procedure allows one to generate a massive dynamic graph of arbitrary size that contains semi-realistic suspicious activities. While meaningful accuracy measures require real data, the structurally congruous synthetic data from AMLSim provides a helpful sandbox for agile research on scalability, explainability, and other requirements.

\textbf{Graph topology:} For this experiment, we used AMLSim to simulate an AML graph with 1M nodes and 9M edges. Each vertex represents an account with attributes including account number, account type (e.g. business type), owner name, and date/time created. Each edge has a transaction ID, amount, and time stamp. The data is sparsely labeled with flagged transactions (i.e. transactions that violate volume and velocity rules) and SARs (full analysis has confirmed suspiciousness).

\textbf{Deep learning task:} Using the SARs and escalated alerts as labels, the task is semi-supervised learning of a predictive model to (a) predict the suspiciousness of a given target node, and (b) identify other potential bad actors in the transaction network via direct or indirect connections to nodes known to be suspicious. To do this, we conduct feature extraction via graph embedding, then convert these features into downstream machine learning tasks.

\textbf{Preliminary experimental results:} The evaluation environment is a workstation with Intel(R) Xeon(R) CPU E5-2697 v4 @ 2.30GHz, total 72 cores (18 cores, 4 sockets) and 1TB DDR4 2400MHz memory. We ran the original, still popularly used deep graph model, GCN~\citep{Kipf2017}, with batched training. GCN took 611 seconds to converge in 32 epochs. Next we ran an improved variant, FastGCN~\citep{Chen2018}, which reduces the training cost through neighborhood sampling. FastGCN took 386 seconds to converge in the same number of epochs. The hyperpameters we used include 128 hidden units, a 0.01 learning rate, and 400 samples in FastGCN. Overall, FastGCN was nearly twice as fast as GCN for this synthetic graph. Projected with a linear scaling, with billion-scale AML graphs (e.g., 1B nodes and 9B edges) the training time would estimate to a few days. We hypothesize the development of high-performance code could improve the training time by additional orders of magnitude for faster turnover given graph dynamism with frequent transaction updates.

 \begin{table}[ht]
   \centering
   \caption{Graph learning on AMLSim data (1M nodes, 9M edges)}
   \label{graph-learning}
   \begin{tabular}{ccccc}
     \textbf{Method} & \textbf{Epochs} & \textbf{Training 	time} \\
     \midrule
     GCN & 32	&	611 seconds \\
     FastGCN	&	32 &	386 seconds \\
   \end{tabular}
   
   \vspace{2mm}
   \scriptsize \textbf{Workstation:} Intel(R) Xeon(R) CPU E5-2697 v4 @ 2.30GHz, 72 cores, 1TB DDR4 2400MHz memory
 \end{table}

\subsection{Optimizing for speed and storage with high-performance computing} 

Deep learning methods for identifying money laundering must be efficient in order to be useful. Financial institutions handle million transactions per second and require timely identification of suspicious activities. The efficiency is explored from both the computation and the memory angle.

\textbf{Sparse computation on graphs:} Systems performing inference using deep neural networks for detecting suspicious activities need the ability to update the graph and node representations sparsely in response to the addition of new transactions. Without dynamic sparse recomputation, a newly added transaction cannot be classified until the entire graph is recomputed. Such a long delay is unacceptable for latency-sensitive applications. Efficient techniques for performing sparse dynamic computation have been developed for a variety of common graph algorithms including PageRank, loopy belief propagation, coordinate descent, co-EM, alternating least squares, singular-value decompositions, and matrix factorizations. For each of these algorithms, sparse dynamic recomputation can significantly enhance performance in practice. 


\textbf{Memory-efficient graph representations:} Storagewise, the parallelism and locality of graph-based deep learning algorithms can benefit from graph reordering in a variety of ways. These algorithms typically perform iterative updates on data associated with a vertex, based on data associated with adjacent vertices. An effective reordering of the graph vertices can improve the data locality of these algorithms, allowing them to use the machine’s memory system more efficiently. In addition, reordering schemes can improve the compressibility of the graph using difference-coding techniques~\citep{Blandford2002}. Finally, many efficient and scalable techniques for processing graphs in parallel, including chromatic scheduling~\citep{Kaler2016}, DAG scheduling~\citep{Hasenplaugh2014,Gruevski2018}, and serial blocking~\citep{Gruevski2018}, benefit from effective graph reorderings, because reordered graph layouts can be leveraged to dramatically reduce synchronization overheads.


\textbf{Preliminary experimental results:} We used the graph compressor, Ligra+ \citep{Shun:2015:SFP:2859848.2860198}, and experimented with the simulated AML graph as well as a few deep learning benchmark graphs used in~\cite{Kipf2017}. Ligra+ is a compression system that achieves reduced space usage while maintaining competitive or even improved performance with respect to running the algorithms on uncompressed graphs on multicore machines. The compression results are reported in Table~\ref{graph-compression}. We see a compression ratio as high as 2x, although larger graphs appear to be more challenging to compress. This finding opens up the opportunity for memory-efficient deep neural network training and inference leveraging compression and reordering techniques.

 \begin{table}[ht]
   \centering
   \caption{Graph compression with Ligra+}
   \label{graph-compression}
   \begin{tabular}{ccc}
     \textbf{Graph}     & \textbf{Size}     &  \textbf{Compressed Size} \\
     \midrule
     Cora     & 59 Kb & 28 Kb \\
     Citeseer & 61 Kb & 29 Kb \\
     Pubmed   & 461 Kb & 252 Kb \\
     AMLSim   & 21 Mb  & 13 Mb \\
     & & \\
   \end{tabular}
 \end{table}


\section{Summary}

In summary, we have motivated the challenge of anti-money laundering not only in terms of legal compliance, but in terms of meaningful societal impact. The human suffering from organized crime such as drug cartel violence and human trafficking is immense, as are the higher order negative effects such as costly penalties for banks and financial exclusion for innocent people on the margins. We have encouraged the technical research community to help financial institutions and law enforcement harness artificial intelligence technologies to fight this problem. We have reviewed current AML methods and provided a first look at deep learning for transaction monitoring and analysis. We have shared preliminary experimental results using synthetic AML data generated from AMLSim. Lastly, we have reviewed some computational considerations involved in implementing graph learning for AML at scale.


%
%

\newpage
\bibliographystyle{plainnat}
\bibliography{reference}

\end{document}